\documentclass[journal]{IEEEtran}

\usepackage{algorithm}
\usepackage[noend]{algpseudocode}
\usepackage{amsmath}
\usepackage{mathrsfs}
\usepackage{dsfont}
\usepackage{graphicx}
\usepackage{epsfig}
\usepackage{epstopdf} 
\usepackage{amssymb}
\usepackage{caption}
\usepackage{subfig}
\usepackage{color}
\usepackage[normalem]{ulem}
\usepackage{soul}

\long\def\comment#1{}

\newfont{\bbb}{msbm10 scaled 700}

\newfont{\bb}{msbm10 scaled 1100}

\newcommand{\gv}{{\bf g}}

\newcommand{\pv}{{\bf p}}

\newcommand{\tv}{{\bf t}}

\newcommand{\Cm}{{\bf C}}

\newcommand{\Gm}{{\bf G}}

\newcommand{\Id}{{\bf I}}

\newcommand{\Qm}{{\bf Q}}

\newcommand{\Sm}{{\bf S}}
\newcommand{\Tm}{{\bf T}}

\newcommand{\Vm}{{\bf V}}

\newcommand{\alphav}{\hbox{\boldmath$\alpha$}}
\newcommand{\betav}{\hbox{\boldmath$\beta$}}
\newcommand{\gammav}{\hbox{\boldmath$\gamma$}}

\newcommand{\zetav}{\hbox{\boldmath$\zeta$}}

\newcommand{\thetav}{\hbox{\boldmath$\theta$}}

\DeclareMathOperator*{\argmin}{arg\,min}

\newcommand{\defeq}{\mathrel{\mathop:}=}

\begin{document}

\title{Object-based High Contrast \\Traveltime Tomography}

\author{Yenting~Lin~\IEEEmembership{Member,~IEEE,}\thanks{Y. Lin is with Google Inc. This work was done during his time at University of Southern California (e-mail:yenting0322@gmail.com).}
  Antonio~Ortega,~\IEEEmembership{Fellow,~IEEE}\thanks{A. Ortega is with the Department
of Electrical Engineering, University of Southern California, Los Angeles, CA 90089, USA (email:antonio.ortega@sipi.usc.edu).} \thanks{Manuscript received ; revised.}
}
\maketitle

\begin{abstract}
  We consider a traveltime tomography problem that involves detection
  of high velocity structures in a homogeneous medium. 
  If we only have limited measurements, this problem becomes an under-determined
  inverse problem and a common approach would be using prior information to guide the reconstruction process.   
  We restrict the possible velocity into discrete values and model it
  as a discrete nonlinear inverse problem. 
  However, typical iterative linearized reconstruction algorithms on  
  grid-spacing model usually have very poor reconstruction results in the presence of high contrast boundaries. 
  The reason is that the travel path bends significantly near the boundary, making 
  it very difficult to infer the travel path and velocity value from
  measured traveltime. 
  To handle this scenario, we propose an object-based approach to
  model high velocity structures by pre-defined convex objects.
  Compared to the typical grid-spacing model, which has variables
  that are proportional to the number of cells, our approach has an advantage that the number of unknown variables
  in the system is proportional to the number of objects, which greatly decreases the problem dimension.  
  We have developed a fast algorithm to provide 
  an estimate of the appearance probability of high velocity
  structures in the region of interest. 
  Simulations show that our method can efficiently
  sample the model parameter space, and provide more robust
  reconstruction results for the scenario where the number of measurements is limited. 
 
\end{abstract}

\begin{IEEEkeywords}
traveltime tomography, Bayesian image reconstruction.
\end{IEEEkeywords}

\IEEEpeerreviewmaketitle

\section{Introduction}

Traveltime tomography aims to reconstruct a velocity
model by using the measured first-arrival time between
transmitters and receivers. 
This technique is widely used to characterize the physical properties of elastic
media, where the heterogeneous structures lead to different travel
velocity. The transmitted signal could be a seismic wave, an acoustic sound wave, 
and even a fluid pressure wave. 
For example, traveltime tomography is applied  
in applications such as seismic geophysical
exploration~\cite{pratt1999seismic}, measuring temperature and flow (in air~\cite{wilson2001overview}
or ocean~\cite{munk1979ocean}) and testing aquifer hydraulic properties \cite{yeh2000hydraulic}. 
However, different from many other tomographic reconstruction problems
(e.g., X-Ray or Positron emission computed tomography), where 
the straight line trajectory assumption is commonly used, in the problems we consider here  
the travel path may bend severely when it encounters a high contrast
velocity abnormality. 
In many geophysical applications, it is common to have heterogeneous
structures that exhibit large differences in velocity between neighboring 
areas, where the scale of heterogeneous structures is much larger than the 
effective wavelength.  
This scenario can be found in different applications: using seismic waves to 
find permeable fracture zones in ground-water 
flow characterization~\cite{national1996rock}, 
monitoring the water/oil saturation in vegetable 
oil bio-remediation projects~\cite{lane2004object} and many others. 
Different from cases where there are only relatively small
changes of velocity in the medium, in high contrast medium the travel path not only bends severely but is also 
dominated by these high velocity structures. 
Thus the straight ray path approximation is no longer a valid assumption, turning reconstruction
of the velocity model into a nonlinear inverse problem.
Bent ray reconstruction methods, based on iteratively finding the travel path and updating the velocity model,
can be used in this context. 
However, in many practical situations the available measured traveltime
data is very sparse, so that these methods perform poorly due to low ray-coverage 
and severe path bending near the boundary of high contrast structures.
For a comprehensive review on this subject, we refer the reader to the overview 
by Berryman~\cite{berryman1991lecture} and to Rawlinson et. al. \cite{rawlinson2008seismic}.

This work is motivated by the problem of flow permeability
characterization in a fractured reservoir
\cite{vasco1997integrating}. Waterflooding, as one of the most widely
used enhanced oil recovery (EOR) techniques, involves 
injecting water in a controlled manner in order to  
offer pressure support that can slowly sweep oil into the 
production wells \cite{welge1952simplified}. 
During this process, the permeability 
of open fractures can be orders of magnitude higher than that
of surrounding tight rocks, providing fast pathways for the flow.
Thus, traveltime through a fracture is much faster than through surrounding rocks. 
The flow properties of the reservoir are dominated by these highly 
`transmissive' zones.
If a fracture is close to both an injection well and a production well, 
most water will flow directly through this fracture and fail to displace oil
in nearby areas.  
This phenomenon is known as ``water cycling'', which significantly
reduces oil recovery efficiency. 
Thus, it is critical to understand the locations of fractures for flow characterization 
and enhancement of oil recovery efficiency.
Lin et. al. \cite{lin2010waterflood, lin2012detecting} investigate
this problem by using the injected water as an input signal and
measuring the change in fluid production, so that the response
time of water injection can be used to provide an estimate of traveltime. 
In this application the physical size of fractures are much larger than the equivalent 
hydraulic wavelength, which makes it a valid traveltime tomography problem.
The major challenge is that the traveltime measurements are limited by the borehole 
locations and the huge velocity change in fractures, which makes it
difficult to reconstruct a high resolution
image to identify fracture locations.
In our previous work \cite{lin2010reconstruction}, we model the
fracture network as lines in a 2D plane and use the total length of
lines as regularization in model reconstruction. 
Here, we extend that approach to model the fracture network by arbitrary
shapes, which gives us more freedom to represent the fractures. 

Many different velocity models have been developed for traveltime
tomography. 
Grid-based models~\cite{Berryman1990notes}, which divide the space volume into small cells and 
assign a constant velocity to each cell, are probably the most popular
ones. These models can represent structures in any degree of detail by increasing
the number of cells. 
As an alternative, object-based models~\cite{lane2004object} use
pre-defined objects instead of fixed size cells to represent the velocity in space. 
Compared to grid-based models, if the pre-defined shape of those
objects is chosen wisely, object-based methods have the advantage of 
representing the spatial distribution of velocity 
with a small number of objects instead of a large number of cells. 
This paper focuses on finding high velocity structures (HVS) 
in a relatively slow homogeneous background, where the background velocity can 
be estimated by geophysical testing. 
For example, a naturally fractured reservoir is usually characterized
by two different types of media: matrix and fracture, where the permeability in 
fractures is much higher than in the matrix medium. 

Only a limited amount of research has addressed the high contrast velocity problem 
in traveltime tomography. 
Bai and Greenhalgh \cite{bai20053} proposed to add irregular cells on
the boundary of reconstructed objects to improve the stability when determining nonlinear ray paths. 
Berryman \cite{berryman1989fermat} used Fermat's principle 
on the reconstruction of velocity model to handle the case where high contrast velocity 
alters the travel path severely.
Zheglova et. al. \cite{zheglova2012level} proposed to reconstruct structure boundaries  by level set inversion. 
All of these approaches use a grid-based model to represent the velocity structure. 
Because the velocity contrast is so high, travel paths change
significantly at the boundaries and most iterative linearized inversion algorithms 
often fail to converge when only very limited measured data is available. 
Also, grid-based models require a
very fine grid to represent the velocity change at the boundaries 
between areas of different velocity.  
A finer grid implies we need to estimate more unknown values 
(the number of parameters to estimate grows linearly with the number of cells), 
so that for some areas  
we cannot determine the travel velocity 
because no travel path passes through the corresponding cells. 
This is the well known  ``lack of ray coverage'' phenomenon. 

As an alternative, in this paper, we extend our previous work
\cite{lin2010reconstruction} to use {\em object-based} models to represent the HVS.
The shape of HVS is approximated by a set of pre-defined convex
objects.  
This approach has two main advantages. 
First, the problem of approximating an arbitrary shape by 
multiple convex objects is well understood. 
Moreover,  we can incorporate prior information about the
structure to define the objects and achieve better model
representations. 
For instance, in the fracture characterization problem 
fractures are known to be well approximated by planes. 
Thus, we can define the fundamental object as a ``rectangular prism'' in 
three dimensional space.  
Compared to other type of objects, such as spheres, only
a small number of rectangles are needed in order to 
approximate the shape of fractures. 
In other words, by choosing the right objects we can use fewer
parameters to model the structure in better detail. 
Fewer model parameters means that there are fewer unknown variables, leading 
to simpler inverse problems. 
Second, the travel path tracking procedure can be simplified by only considering 
the shortest path between objects instead of cells in spatial domain. 
Compared to the above methods mentioned, it avoids the
``lack of path coverage'' problem arising in grid-based models, 
which leads to  a more stable solution to the problem. 

To recover the velocity model from the measured traveltime, we need
to solve a nonlinear inverse problem. 
The challenge in all inverse problems is that the solution, 
in this case the estimated velocity model, may not be unique when 
limited measurements are available. 
One popular approach to handle the non-uniqueness issue is to 
apply regularization to favor certain properties in 
the model \cite{engl1996regularization}. 
The regularization approach can be viewed as model selection: 
it will lead to a solution that balance data-fitting and model-penalty.         
However, it is not trivial to choose a suitable weight for model-penalty and
this usually requires cross validation in order to avoid over-fitting \cite{ng1997preventing}. 

An alternative approach is to estimate 
the probability of different models in the model parameter space according to the data-fitting \cite{tarantola1987inverse}. 
This gives a full description of the relative probabilities of different models, 
so that we can explicitly consider all likely solutions.    
However, generating samples and estimating the probability density 
is very challenging and time consuming for a high dimensional 
model space \cite{mackay2003information}.  

In this paper, we choose the second approach and use an efficient
Hybrid Monte Carlo (HMC) sampling on the probability 
density function. 
In order to visualize the result, we introduce the ``appearance probability map'' 
which indicates where the high velocity objects are more 
likely to appear in the spatial domain. 
To the best of our knowledge, we are the first to 
use an object-based approach to solve a 
 high contrast, discrete velocity tomography problem.
Our proposed algorithm uses the HVS properties to simplify the travel path
finding step, which makes the HMC sampling process much more efficient.   

The rest of the paper is organized as follows.   
In Section \ref{sec:object_based_velocity} 
we define the object-based model to represent the different velocity
structures. 
This is followed in Section \ref{sec:problem_formulation} by the introduction of the forward operator 
and the mathematical formulation for the travel path
finding problem. 
Then we give an overview of our proposed
algorithm for sampling the probability distribution function in Section \ref{sec:inversion}. 
Simulation results are given in Section \ref{sec:sim_results} and
conclusions are presented in Section \ref{sec:conclusion}. 

\section{Object based model}
\label{sec:object_based_velocity}

In computer graphics, it is well established that we can approximate arbitrary structures in any
level of details by increasing the number of  
objects \cite{agarwal1998surface}. 
Triangles, quadrilaterals or other simple convex polygons are very
popular choices of fundamental objects in geometric modeling. 
In this work, we do not restrict the fundamental object to be any
specific type of geometrical shape. Instead, we use an abstract ``convex'' object, where the
actual type of shape can be defined as a parameter. 

\begin{figure}[htb]
  \centering
    \begin{subfloat}[]
{\includegraphics[width=0.45\linewidth]{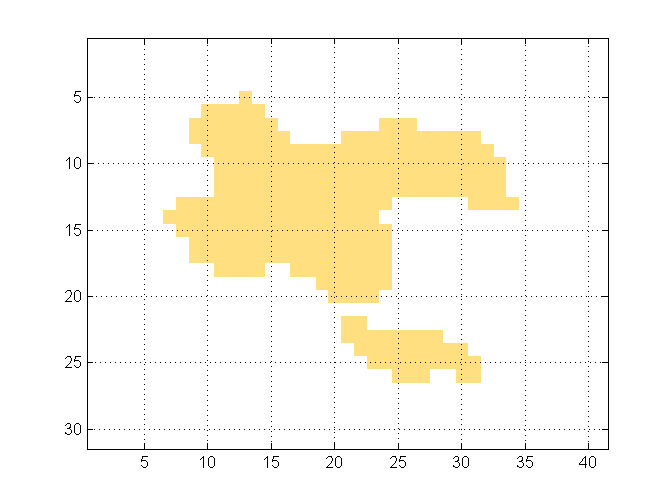}}
    \end{subfloat}
\begin{subfloat}[]
   {\includegraphics[width=0.45\linewidth]{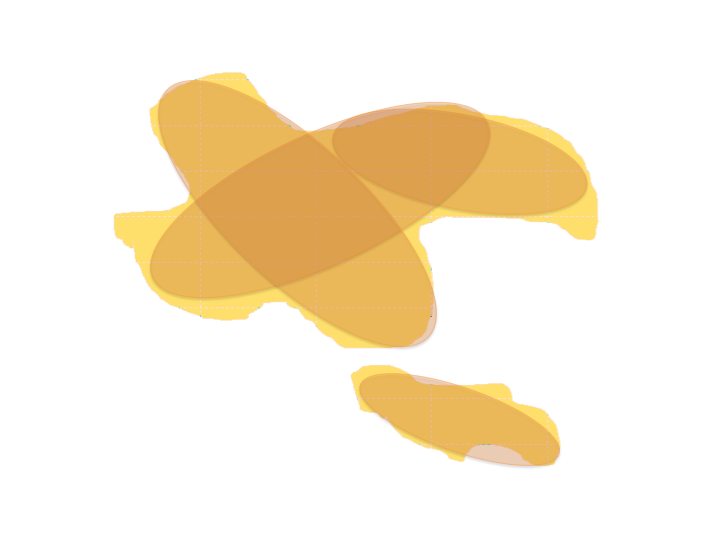} }    
\end{subfloat}
  \caption{(a)  Grid based model:  the HVS structure is approximated
    by cells. (b) Object based model:  the structure is approximated by objects.}
  \label{fig:grid_object_model}
\end{figure}

For example, if we choose ellipse as the type for the $i$-th object, the remaining parameters will be 
center location, orientation, major and minor axis.    
This leads to a vector representation of parameters 
$\thetav_{i} = \{\omega_{i}, x_{i}, y_{i}, \psi_{i}, a_{i}, b_{i} \}$ 
where $\omega_{i}$ represents the type of object.  
Denote $| {\thetav}_{i} | $ the volume inside the $i$-th object. 
Different from the typical geometric modeling, in our model we allow the objects to
overlap with each other (see Fig. \ref{fig:grid_object_model}).  

Next, let $\{ \thetav_{1}, \thetav_{2}, \dots, \thetav_{N} \}$ be the
set of $N$ high velocity objects, we
denote $v_{h}$ the velocity in the homogeneous background and $v_{i}$  
the velocity for object $\thetav_{i}$. 
Because we may have several objects overlapping with each other, the
velocity in the overlapping volume has to be defined carefully. 
In this case, we define the velocity at location $(x,y)$ 
as the maximum velocity among all objects that include $(x,y)$:
\begingroup\makeatletter\def\f@size{9}\check@mathfonts
\begin{equation}
 V(x,y)= 
\left\{
\begin{array}{cc} 
\displaystyle {\max_{i}} ~v_{i}, &  (x,y) \in |{\thetav}_{i}| \\
v_{h},  & (x,y) \not\in  \overset{N}{\underset{i=1}{\cup}} |{\thetav}_{j}|
\end{array}
\right.
\end{equation}
\endgroup
Obviously the spatial velocity distribution is implicitly determined by the
location of objects.
We use the notation $V(\thetav_{1}, \dots, \thetav_{N})$ to indicate the velocity distribution 
when we have $N$ objects with parameters $\thetav_{1}, \dots,
\thetav_{N}$ in the model. 
Following the same notation, we use $d(\thetav_{i}, \thetav_{j})$ to represent the distance
between two objects, which is given by:
\begingroup\makeatletter\def\f@size{9}\check@mathfonts
\begin{equation}
\label{eqn::define_shortest_dist}
d(\thetav_{i}, \thetav_{j}) = \min_{\boldsymbol{\mu, \nu}} \|
\boldsymbol{\mu} - \boldsymbol{ \nu} \|_2 , 
~ \boldsymbol{\mu} \in |{\thetav}_{i}|, ~ \boldsymbol{\nu} \in |{\thetav}_{j}| 
\end{equation} 
\endgroup
and the corresponding path is denoted by $\vec{P}({\thetav_{i}, \thetav_{j}})$. 
Obviously, if $\thetav_{i}$ overlaps with $\thetav_{j}$ then $d(\thetav_{i}, \thetav_{j}) = 0$. 
The same notation can be used for the distance between a point and an object, 
or between two points, e.g., $d(\alphav, \betav)$ and $\vec{P}({\alphav, \betav})$ 
are the distance and path between points ${\alphav}$ and ${\betav}$, respectively.

\section{Forward step}
\label{sec:problem_formulation}

In this paper, we address the problem of identifying high velocity structures
in homogeneous background with sparse data. 
To solve this problem, we need a ``forward'' model to compute the
traveltime for a given velocity model. 
Then, in the inverse step, a velocity model is estimated by minimizing the
mismatch between predicted and measured traveltime data. 

Given the velocity model as input, the forward model will compute the
corresponding traveltime between transmitter and receiver. 
We use the mathematical formulation proposed 
in \cite{berryman1991lecture} to calculate the traveltime, where the
travel path is defined as the direction of wave-front 
propagation. 
Based on Fermat's principle, the actual travel path is the one with
minimum cost time from all possible paths connecting the transmitter and the receiver.
The time cost of an arbitrary path $P$ connecting two points $( \alphav, \betav )$
based on velocity model $V$ is defined by 
the path integral:
\begingroup\makeatletter\def\f@size{9}\check@mathfonts
\begin{eqnarray}
\tau^{P}(V, \alphav, \betav) = \int_{P} \frac{1}{V(x,y)} dl^{P},
 ~\{P_{sart}, P_{end}\} = \{\alphav, \betav \}.
\end{eqnarray}
\endgroup 
The travel path, $P^*$, is defined as the path with minimum time cost $\tau^{*}$.
Therefore, we can also define the travel time $\tau^{*}$ between two points 
$\{ \alphav, \betav \}$ as: 
\begingroup\makeatletter\def\f@size{9}\check@mathfonts
\begin{eqnarray}
\label{eqn:travel_time_equation}
\tau^{*}(V, \alphav, \betav) & = & \min_{P} \tau^{P}(V, \alphav,
\betav)  \nonumber \\
& = & \min_{P} \int_{P} \frac{1}{V(x,y)} \, dl^{P} .  
\end{eqnarray}
\endgroup
The travel path $P^*$ will be: 
\begingroup\makeatletter\def\f@size{9}\check@mathfonts
\begin{equation}
\label{eqn::travel_path_equation}
P^{*}(V, \alphav, \betav) = \argmin_{P} ~\tau^{P}(V, \alphav, \betav) .
\end{equation}
\endgroup
Finding the analytic solution for the traveltime $\tau^*$ and travel path $P^{*}$ is a 
classical problem in calculus of variations \cite{courant1962methods}.
Most approaches, including shotgun ray-tracing or
fast marching \cite{beydoun1987paraxial,vidale1990finite,sethian19993}
tend to be very computationally intensive due to the frequent update of
traveltime in each cell.

\subsection{Fast travel time/path finding}

The object-based model we proposed in Section 
\ref{sec:object_based_velocity} is a generic model. 
Each object can have arbitrary velocity, and the model has the 
ability to present any velocity distribution. In this paper we use 
Markov Chain Monte Carlo (MCMC) sampling for reconstruction. 
Typically Monte Carlo sampling is considered to be too computational expensive for traveltime tomography.
The reason is that in each iteration we would need to calculate the traveltime for a new model, and then run a large number iterations to sample the probability space.
Thus, it is critical to find a way to calculate the traveltime efficiently.

Since we consider the high contrast velocity case, 
for simplicity here we assume that all objects have the
same velocity, which is much higher than the background velocity, 
i.e., $ v_{1} = v_{2} = \dots =v_{N}$ and $v_{i} \gg v_{h} $. 
This assumption will not be valid for smooth velocity variation. But for the 
application we consider, fractured reservoir characterization, the fracture 
permeability is several orders of magnitude higher 
than that of the host rocks.     
With the high velocity contrast assumption, the time spent passing through an object can be ignored, which  
provides a way to find the fastest travel path  by considering the path between objects recursively, thus 
greatly reducing the computational complexity. 
We will explain this concept next. 

It is well known that the travel path is a straight line in homogeneous medium. 
We start by considering a simple scenario, where there is only one object in the
model. 
Obviously, there are two possible cases for the travel path, i.e., the
travel path may or may not traverse the object.  
If the path does not traverse the object, the whole path will belong to
the homogeneous background, which implies that it must be a straight ray
connecting transmitter and receiver.  
Otherwise, if the path passes through the object, it will be a combination of
three parts, corresponding to the paths i) from transmitter to
object, ii) inside object and iii) from object to receiver. 
Each part traverses a constant velocity medium, thus, the
travel path must be a combination of line segments. 

With the high contrast velocity assumption, the travel time inside the volume
$|{\thetav_{1}}|$ is negligible. 
Thus, the travel path/time through the object must be 
the ``shortest'' path from transmitter to $|{\thetav_{1}}|$, a path inside
$|{\thetav_{1}}|$ and the ``shortest'' path from $|{\thetav_{1}}|$
to the receiver, which are all straight lines. 
We use $\vec{P}(\alphav, \gammav)$, $\vec{P}(\gammav,
\zetav)$ and $\vec{P}(\gammav, \betav)$ to denote these lines, 
where $\gammav, \zetav$ are the closest points to transmitter
$\alphav$, receiver $\betav$ in
$|\thetav_{1}|$, which are defined as:
\begingroup\makeatletter\def\f@size{9}\check@mathfonts
\begin{align}
  \label{eq:def_short}
  \gammav = \argmin_{\gammav} \| \alphav - \gammav \|, & ~\zetav = \argmin_{\zetav} \|
  \betav - \zetav \| \\ 
   & \text{where} ~ \{ \gammav, \zetav \} \in |\thetav_{1}| . \nonumber  
\end{align}
\endgroup

This travel path can be denoted as $\alphav-\thetav_{1}-\betav$ which
describes the order in which the objects are traversed. 
If we assume that the traveltime inside the object is 
negligible, the traveltime function in the one-object case becomes 
\begingroup\makeatletter\def\f@size{9}\check@mathfonts
\begin{equation}
\label{eqn:travel_time_extreme_one_obj}
\tau^{*}(V(\thetav_{1}),\alphav, \betav) = 
\min \left\{
\begin{array}{c}
1/v_{h} \cdot {d(\alphav, \betav)} \\
1/v_{h} \cdot ( ~d(\alphav, \thetav_{1}) + d(\thetav_{1}, \betav) ~) 
\end{array}
\right.
\end{equation}
\endgroup

Next, we consider the two object case $N=2$. 
Obviously, there are three possible cases for the
travel path, namely, i) not traversing any object, ii) traversing
exactly one object, or iii) traversing both objects. 
From (\ref{eqn:travel_time_extreme_one_obj}) we 
know how to compute the travel path/time in the cases when a single
object is traversed. 
Thus,  we only need to analyze how to compute the paths  
which pass through both objects, $|{\thetav}_{1}|$ and
$|{\thetav}_{2}|$, in different orders, 
and then compare them with the previous results. 

We notice that the path $\alphav -  \thetav_{1} - \thetav_{2} - \betav$ includes 
the shortest paths from $\alphav$ to object $\thetav_{1}$, 
from object $\thetav_{1}$ to object $\thetav_{2}$, and from object $\thetav_{2}$ to $\betav$.
Similar to  (\ref{eqn:travel_time_extreme_one_obj}), the corresponding travel time will be 
$1/v_{h} \cdot ( ~d(\alphav, \thetav_{1}) +  d(\thetav_{1},
\thetav_{2}) +  d(\thetav_{2}, \betav) ~)$, where $d(\alphav,
\thetav_{1})$ has been calculated in the previous step in order to
evaluate the path $\alphav - \thetav_{1} - \betav$. 
Thus, $d(\thetav_{1}, \thetav_{2})$ and $d(\thetav_{2}, \betav)$ are
the only new terms to be calculated.
A similar result also holds for $\alphav - \thetav_{2} - \thetav_{1} -
\betav$, where only $d(\alphav, \thetav_{2})$ and $d(\thetav_{2},
\thetav_{1})$ would be needed. 
Note that $d(\thetav_{2}, \thetav_{1}) = d(\thetav_{1},
\thetav_{2})$ so that this 
distance does not need to be recomputed. 

In summary, given two objects
in the velocity model
the traveltime can be computed as (again the traveltime inside objects is considered negligible):
\begingroup\makeatletter\def\f@size{9}\check@mathfonts
\begin{align}
\label{eqn:path_tracking_simple}
\tau^{*}(V(\thetav_{1}, \thetav_{2}),\alphav, \betav)  & =  \nonumber \\
1/v_{h} \cdot 
\min & \left\{
\begin{array}{c}
  d(\alphav, \betav) \\
  d(\alphav, \thetav_{1}) + d(\thetav_{1}, \betav)  \\
  d(\alphav, \thetav_{2}) + d(\thetav_{2}, \betav)  \\
  d(\alphav, \thetav_{1}) + d(\thetav_{1}, \thetav_{2})   + d (\thetav_{2}, \betav) \\ 
  d(\alphav, \thetav_{2}) + d(\thetav_{2}, \thetav_{1})   + d (\thetav_{1}, \betav)
\end{array}
\right.
\ldotp
\end{align} 
\endgroup

Note that we need to compute and compare $5$ different possible
paths in the 2-object case. But two of them $\{ d(\alphav, \thetav_{1}) + d(\thetav_{1}, \thetav_{2}) + d(\thetav_{2}, \betav), ~ d(\alphav, \thetav_{2}) + d (\thetav_{2}, \betav) \}$ are actually
trying to find the fastest path from $\alphav$ to object
$\thetav_{2}$, then travel from $\thetav_{2}$ to $\betav$. Thus, Equation
\eqref{eqn:path_tracking_simple} can be reformulated as 
\begingroup\makeatletter\def\f@size{9}\check@mathfonts
\begin{align}
  \label{eq:path_new_format}
\tau^{*}(V(\thetav_{1}, \thetav_{2}),& \alphav, \betav)  =  \nonumber \\
\min & \left\{
 \begin{array}{c}
  1/v_{h} \cdot  d(\alphav, \betav) \\
  \tau^{*}(V(\thetav_{1}, \thetav_{2}),\alphav, \thetav_{1}) + 1/v_{h}
  \cdot d(\thetav_{1}, \betav)  \\
  \tau^{*}(V(\thetav_{1}, \thetav_{2}),\alphav, \thetav_{2}) + 1/v_{h}
  \cdot d(\thetav_{2}, \betav)  \\
\end{array}
\right.
\ldotp  
\end{align}
\endgroup
where 
\begingroup\makeatletter\def\f@size{9}\check@mathfonts
\begin{equation}
  \label{eq:path_explain_propageate}
\tau^{*}(V(\thetav_{1}, \thetav_{2}),\alphav, \thetav_{1})  =  
1/v_{h} \cdot \min \left\{
\begin{array}{c}
   d(\alphav, \thetav_{1}) \\
  d(\alphav, \thetav_{2}) + d(\thetav_{2}, \thetav_{1})
\end{array}
\right.
\ldotp    
\end{equation}
\endgroup

It is obvious that if we have $2$ objects, the
traveltime can be computed by comparing the fastest path from
transmitter to each object $\thetav_{i}$. 
Equation \eqref{eq:path_new_format} can be extended to the case with
$N$ objects:
\begingroup\makeatletter\def\f@size{9}\check@mathfonts 
\begin{align}
  \label{eq:traveltime_all}
& \tau^{*}(V(\thetav_{1}, \dots, \thetav_{N}), \alphav, \betav)  =  \nonumber \\
& \min \left\{
\begin{array}{c}
  1/v_{h} \cdot  d(\alphav, \betav) \\
  \tau^{*}(V(\thetav_{1}, \dots, \thetav_{N}),\alphav, \thetav_{1}) + 1/v_{h}
  \cdot d(\thetav_{1}, \betav)  \\
  \vdots \\
  \tau^{*}(V(\thetav_{1}, \dots, \thetav_{N}),\alphav, \thetav_{N}) + 1/v_{h}
  \cdot d(\thetav_{N}, \betav)  \\
\end{array}
\right.
\ldotp  
\end{align}
\endgroup 
and 
\begingroup\makeatletter\def\f@size{9}\check@mathfonts 
\begin{align}
  \label{eq:traveltime_append}
& \tau^{*}(V(\thetav_{1}, \dots, \thetav_{N}),\alphav, \thetav_{i})  =  \nonumber \\
& \min \left\{
\begin{array}{c}
   1/v_{h} \cdot  d(\alphav, \thetav_{i}) \\
  \tau^{*}(V(\thetav_{1}, \dots, \thetav_{N}),\alphav, \thetav_{1})  +
  1/v_{h} \cdot d(\thetav_{1}, \thetav_{i}) \\
\vdots  \\
  \tau^{*}(V(\thetav_{1}, \dots, \thetav_{N}),\alphav, \thetav_{N})  +   1/v_{h} \cdot d(\thetav_{N}, \thetav_{i})
\end{array}
\right.
\ldotp      
\end{align}
\endgroup

Note that in equation \eqref{eq:traveltime_append}, the calculations of traveltime seem to be
correlated with each other and hard to compute. 
However, we notice that the distances between objects are always
``non-negative'', which provides us an opportunity to simplify the \textit{min}
operation in the calculation.

To show how this property can dramatically simplify the
calculation, we start the calculation by finding the ``closest'' object $\thetav_{k_{1}}$ from $\alphav$.
This can be found immediately by comparing the distance from transmitter
to all objects: 
\begin{equation}
  \label{eq:explain_dijk}
 \thetav_{k_{1}} = \underset{\thetav_{i}}{\argmin}  ~d(\alphav, \thetav_{i})  
\end{equation}
and the traveltime will be:
\begin{equation}
  \label{eq:explain_dijk_2}
 \tau^{*}(V(\thetav_{1}, \dots, \thetav_{N}),\alphav, \thetav_{k_{1}})
= 1/v_{h} \cdot d(\alphav, \thetav_{k_{1}})   
\end{equation}

Why does the travel path to the closest object only contain the direct
path from transmitter? It can be easily proved by
the following statements. 
Assume that $\thetav_{k_{1}}$ is the closest object, and the travel path 
is not the direct path, i.e., the 
travel time/path is a combination of the path through another
object $\thetav_{j}$. In other words, $\tau^{*}(V(\thetav_{1}, \dots, \thetav_{N}),\alphav, \thetav_{k_{1}}) = \tau^{*}(V(\thetav_{1}, \dots, \thetav_{N}),\alphav, \thetav_{j}) + 1/v_{h} \cdot d(\thetav_{k_{1}},
\thetav_{j})$.  But the distance between objects is always greater
than zero, $d(\thetav_{k_{1}}, \thetav_{j}) \geq 0$, therefore 
$\tau^{*}(V(\thetav_{1}, \dots, \thetav_{N}),\alphav, \thetav_{k_{1}}) \geq \tau^{*}(V(\thetav_{1}, \dots, \thetav_{N}),\alphav, \thetav_{j})$. 
This implies that another
object $\thetav_{j}$ has shorter traveltime than $\thetav_{k_{1}}$, which contradicts the
assumption and completes our proof. 

With the same concept, we can find the second ``closest'' (shortest
traveltime) object by considering the path traverses
through $\thetav_{k_{1}}$. The cost time to each object $\thetav_{i}$ through $\thetav_{k_{1}}$ will be:  
\begingroup\makeatletter\def\f@size{9}\check@mathfonts 
\begin{align}
  \label{eq:explain_dijk_2b}
\tau(V(\thetav_{1}, & \dots, \thetav_{N}), \alphav, \thetav_{i}) =  \nonumber \\
\min & \left\{
\begin{array}{c}
   1/v_{h} \cdot  d(\alphav, \thetav_{i}) \\
  \tau^{*}(V(\thetav_{1}, \dots, \thetav_{N}),\alphav, \thetav_{k_{1}})  +
  1/v_{h} \cdot d(\thetav_{k_{1}}, \thetav_{i}) \\
\end{array}
\right.
\ldotp      
\end{align}
\endgroup
Then the second ``closest'' object $\thetav_{k_{2}}$ can be selected with the
minimum of cost time among these objects.  
\begingroup\makeatletter\def\f@size{9}\check@mathfonts 
\begin{equation}
  \label{eq:explain_dijk_3}
 \tau^{*}(V(\thetav_{1}, \dots, \thetav_{N}),\alphav, \thetav_{k_{2}})
= 
 \displaystyle{\min_{i, i \neq k_{1}}} \tau(V(\thetav_{1}, \dots, \thetav_{N}),\alphav,
\thetav_{i}) 
\end{equation}
\endgroup

Following this procedure, after $n$ iterations we can find the $n$-th
``closest'' object. This process is very similar to the fast marching method to solve the
Eikonal problem, which never backtracks over previously evaluated
grid points. In the next section, we will use this concept and introduce an efficient algorithm to
compute the travel path. 

\subsection{Dijkstra path finding}

\begin{figure}[htb]
  \centering
\begin{subfloat}[]{\includegraphics[width=0.4\linewidth]{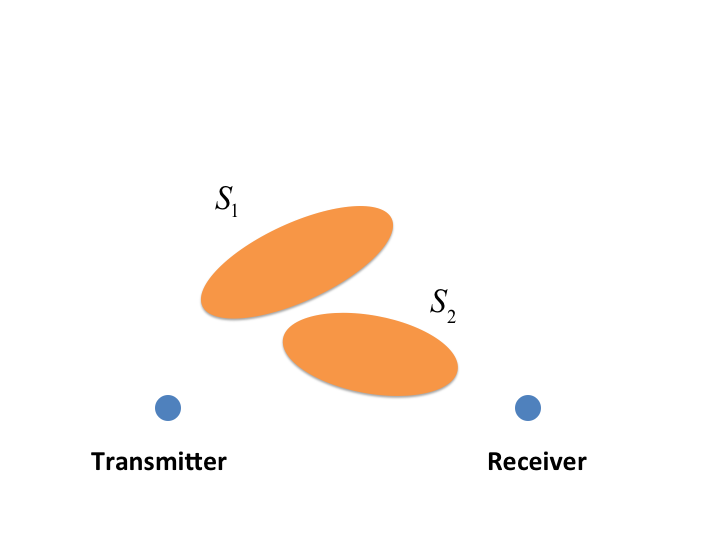}
\label{fig:two_possible_path}
}
\end{subfloat}
\begin{subfloat}[]{
  \includegraphics[width=0.4\linewidth]{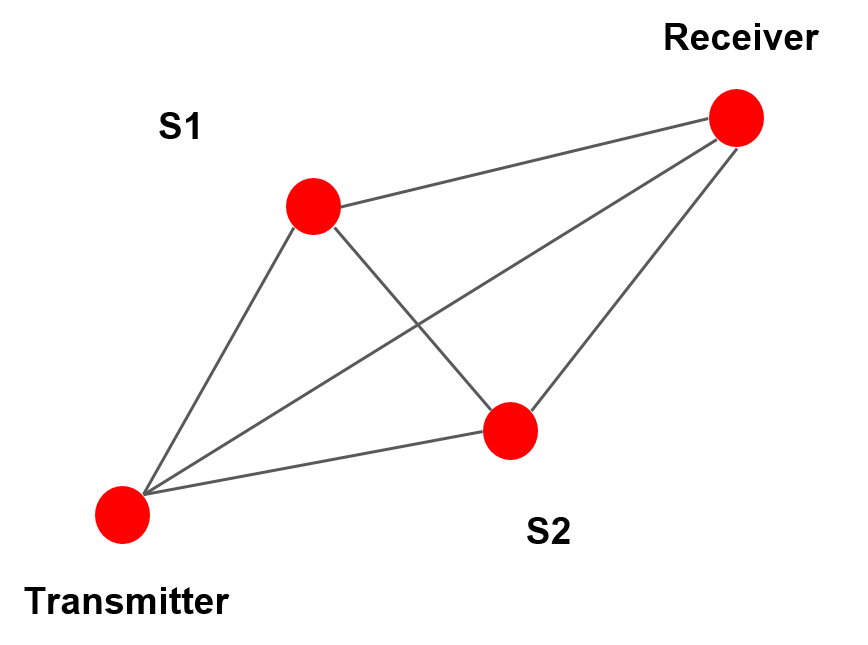}
  \label{fig:convert_dijkstra}
}
\end{subfloat}
\caption{Example of graph representation of an object model. (a)
  Geometrical locations of the objects
  (b)  Graph model, where the edge weight
  is the geometrical distance between objects.}
\label{fig::dijk_path_graph}
\end{figure}

Note that in ~\eqref{eq:path_new_format}, as well as in more general 
cases with more objects, the traveltime is
obtained by comparing different possible paths.  
Assuming that we would like to find the traveltime from the
source to the destination, the fastest path may traverse one or more
objects; which objects and in which order they are traversed is
not known. 
From the previous discussion, we know the fastest path from
transmitter to receiver can be found by iteratively finding the
``closest'' object. 
To solve this problem systemically, we construct a graph $G=(v,e)$ where each node
represents an object, and additional source and destination nodes are
defined for the transmitter and receiver.  
This graph is fully connected with N+2 nodes, and the edge weight between the
nodes is the distance between the corresponding objects defined in (\ref{eqn::define_shortest_dist})
(sources, destinations or objects, see Fig. \ref{fig::dijk_path_graph}). 

\begin{figure}[htb]
     \centering
\begin{subfloat}[]{
   \includegraphics[width=0.28\linewidth]{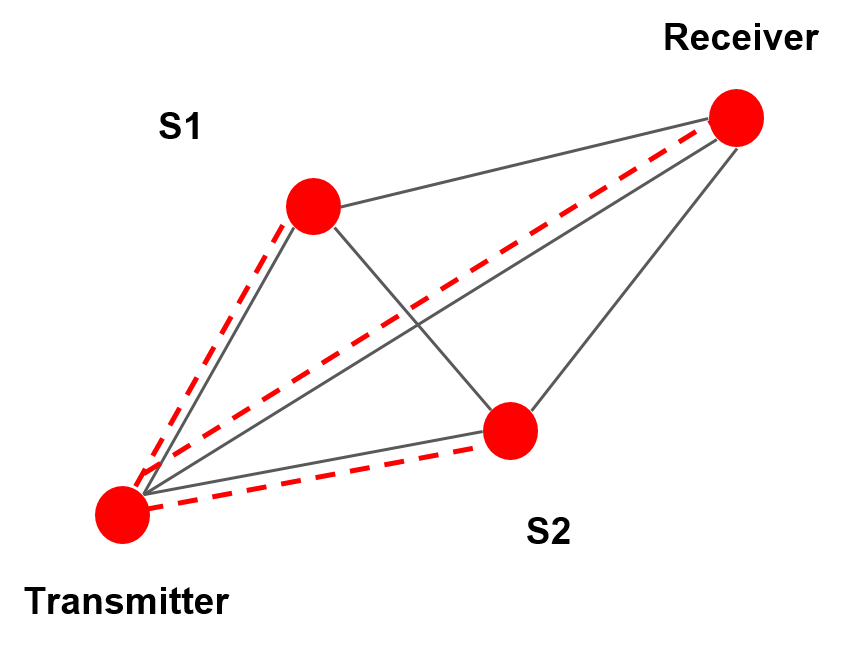} 
}
\end{subfloat}
\begin{subfloat}[]{
    \includegraphics[width=0.28\linewidth]{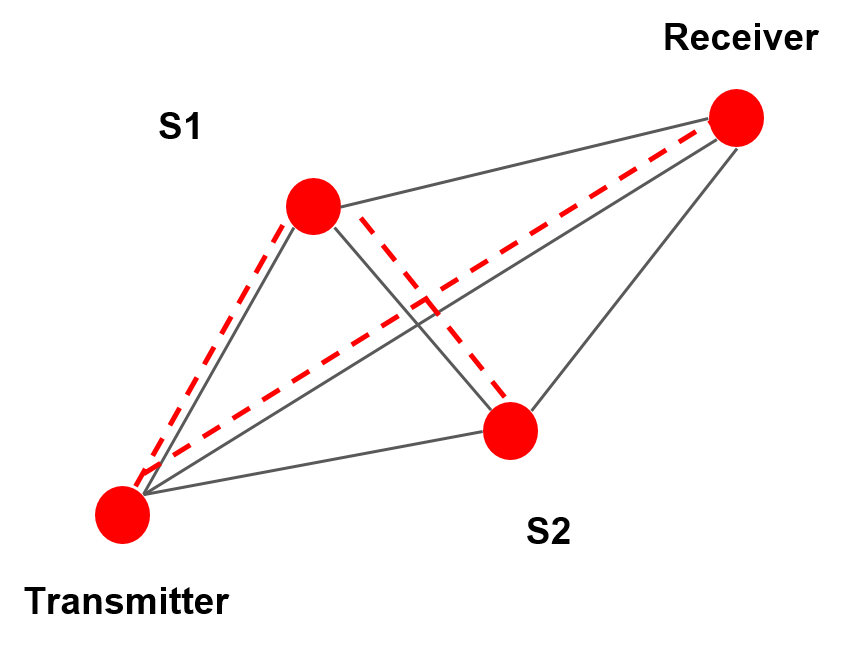} 
}
\end{subfloat}
\begin{subfloat}[]{
    \includegraphics[width=0.28\linewidth]{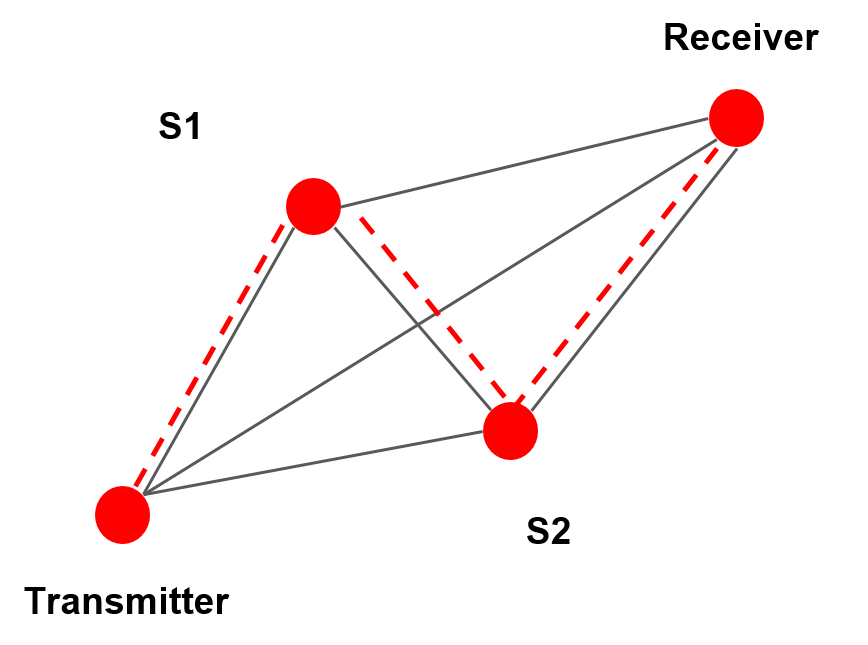} 
}
\end{subfloat}
  \caption{Example of the Dijkstra algorithm to find the travel
    path. (a) After initialization, the cost to each node will be the
    direct path. (b) The cost and path to $S_{2}$ 
    are updated by the path through $S_{1}$. (c) The final result. The
    traveltime is determined by the path through $S_{1}$ and $S_{2}$.  }
\label{fig:dijk_process}
\end{figure}

After constructing the graph and computing the edge weight 
(i.e., the pairwise distance between objects), we apply the
Dijkstra algorithm to find the shortest distance from
source (transmitter) node to destination (receiver) node. 
The Dijkstra algorithm iteratively construct a shortest-path tree from the
source node to every other node in the graph.
We demonstrate the construction of shortest-path tree and update of distance metric by considering an example
with two objects. The geometrical location of objects and the corresponding graph are shown in
Fig. \ref{fig::dijk_path_graph}, while the distance metric updates 
are demonstrated in Fig. \ref{fig:dijk_process}. 
In the initial stage, the traveltime from source $\alpha$ to all other
nodes is unknown, thus the metric on all other nodes is set to infinity.  
Then, in the first step the metric is updated by the
direct path from $\alpha$ to all other nodes, shown in
Fig. \ref{fig:dijk_process}(a). 
Node $S_{1}$ is selected as the ``closest'' object, and we update the
metric for neighbors of $S_{1}$ by comparing the tentative
distance and the recorded distance. 
For instance, node $S_{1}$ has distance metric $3$ and the
edge between $S_{1}, S_{2}$ is equal to $1$, thus the ``tentative''
path to $S_{2}$ has cost $v(S_{1}) + e(S_{1}, S_{2}) = 4$. 
Then we compare this new tentative path with the original cost and
choose the minimum.   
Thus, the metric in node $S_{2}$ is updated to $4$ and node $\beta$ is
updated to $8$.  
By iteratively selecting the closest node, the fastest path to $\beta$
is eventually found with the path $\alpha-S_{1}-S_{2}-\beta$. 

\begin{algorithm}[htb]
\caption{Dijkstra algorithm for path tracking}
\label{alg:dijkstra}
\begin{algorithmic}
\State dist[$s$] $\leftarrow 0$ \Comment{\small distance to source is zero}
 \For {all {$v \in \Vm \setminus s$}}         \Comment{\small initialization}                         
 \State          dist[$v$] $\leftarrow \infty$  
\EndFor
\State       $\Sm = \emptyset$         \Comment{\small initially the set of visited vertices is empty} 
\State      $\Qm \leftarrow \Gm$     

\While {$\Qm \neq \emptyset$}                      
\State          $u \defeq v \in \Qm$ with minimum dist$[v]$ 
\State          $\Sm \leftarrow \Sm \cup u$ 
\State          $\Qm \leftarrow \Qm \setminus u$ 
\For { all $v \in$  neighbor[$u$] and $v \in \Qm$}                  
\If              {dist$[u] + e(u, v) \leq$ dist$[v]$}  
\State                  dist$[v]$ $\leftarrow$ dist$[u] + e(u, v)$  
\EndIf
\EndFor
\EndWhile
\end{algorithmic}
\end{algorithm}

\begin{figure}[htb]
  \centering
    \begin{subfloat}[]{
\includegraphics[width=
    0.4\linewidth]{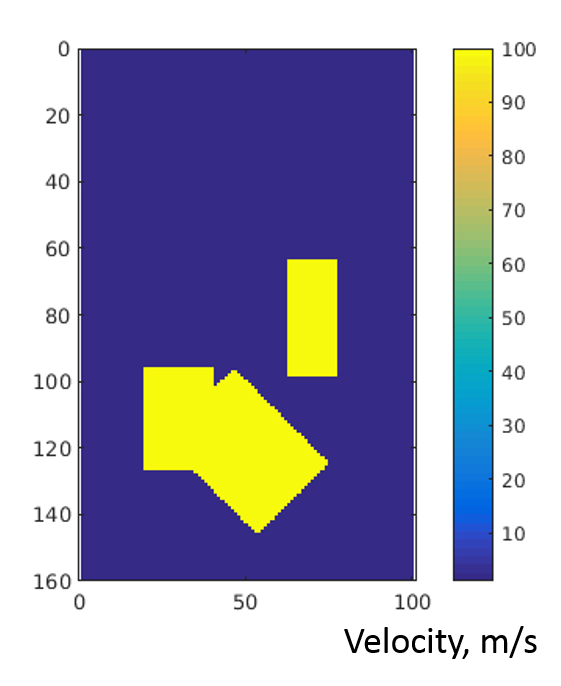}
    }
\end{subfloat}
    \begin{subfloat}[]{
\includegraphics[width=
    0.4\linewidth]{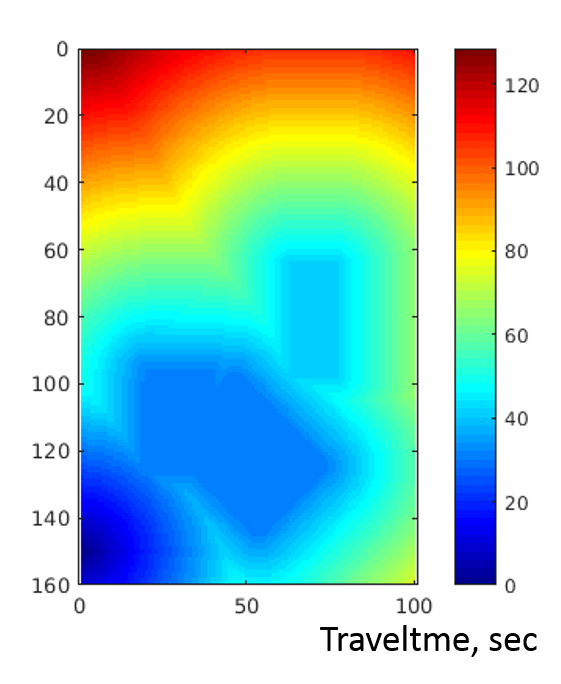}
    }
\end{subfloat}
\begin{subfloat}[]{
   \includegraphics[width=
   0.4\linewidth]{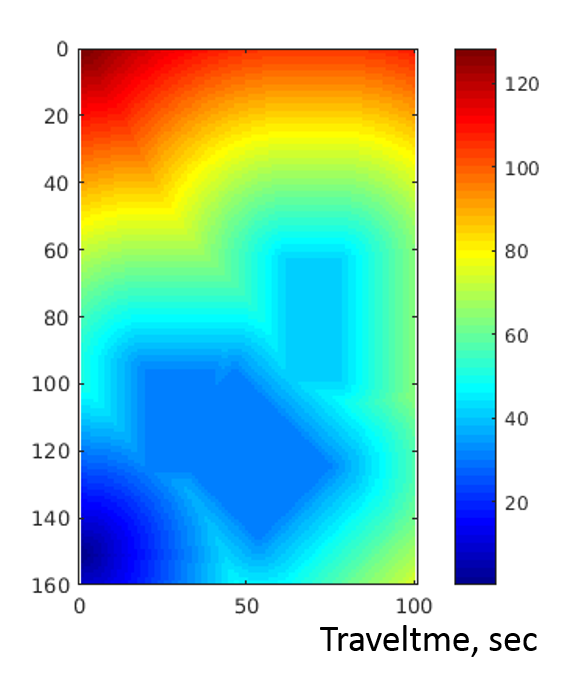}     
} 
\end{subfloat}
   \begin{subfloat}[]{
\includegraphics[width=
   0.4\linewidth]{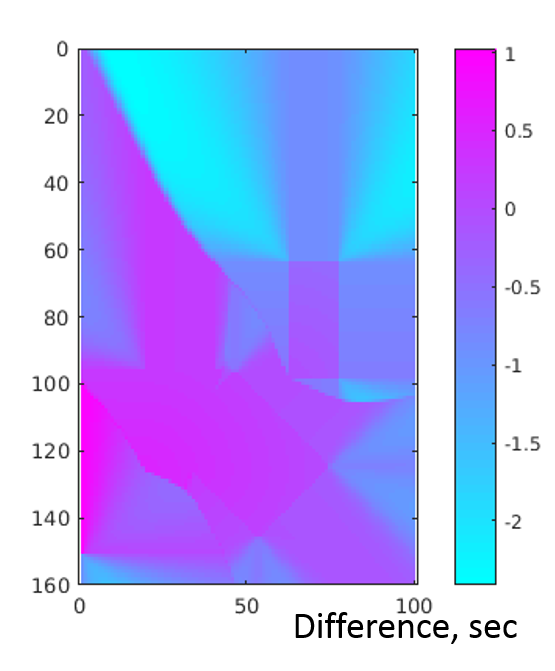}
   }
\end{subfloat}
  \caption{Distance map: the traveltime from transmitter
      in $(1, 150)$ (a) Base velocity model, where the background
      velocity is $1 ~m/s$ and object is $100 ~m/s$ (b) Result
      obtained with our approach (c) Result from fast marching method (d) Difference between two approaches. Note most of the
    differences are within $5\%$. }
  \label{fig:distance_map}
\end{figure}

To validate our method, we build a 2D velocity model with size $100 ~\text{m} \times
160 ~\text{m}$ and velocity equal to $1 ~m/s$ in the background and
$100 ~m/s$ in the objects. 
The transmitter is located in the lower left corner of the model and we calculate the 
``distance map'', representing the 
traveltime from the transmitter to each point.
In Fig. \ref{fig:distance_map}, we show the results from our approach
and the well know fast marching method \cite{sethian1996fast, kroon2009accurate} which uses rectangular cells
with size $1 \text{m} \times 1 \text{m}$. 
The calculated distance map from our approach is very close to that
obtained from fast marching method, but our method has much faster
computational time. 
The reason is that both methods convert the travel path finding 
into a shortest path problem in graph. 
However, instead of using the rectangular cells to represent the velocity
model, our approach uses ``objects'' to represent the velocity
model which reduces the number of nodes in the
graph from $100 \times 100$ (number of cells) to $4$. 
When running the Dijkstra algorithm to find the shortest path on the
graph (while the simplest implementation has computational complexity 
$O(|V|^2)$ and memory requirement $O(|V|)$), the cost time drops
from $O(10,000^2)$ to $O(4^2)$.  

\section{Inverse Problem Solution}
\label{sec:inversion}
 
In the forward step we just presented, we can predict the traveltime
when the velocity model is given.
Next, for the inverse step the goal is to
estimate the velocity model when the traveltime data is observed.  
With limited measured data, this becomes an ill-posed inverse problem and   
the solution may not be unique. 
Since multiple solutions may exist, rather than selecting a single model as the
output of our algorithm, we formulate it as a  
statistical inference problem and estimate the probability
distribution in the velocity model parameter space. 
We start this section by introducing some notations.   

\subsection{Notation}

The input data is obtained by measuring the traveltime between the set of transmitters 
$\mathcal{A}= \{ \alphav_{1}, \dots, \alphav_{p} \}$ 
and receivers $\mathcal{B} = \{ \betav_{1}, \dots, \betav_{q} \}$. 
We denote the measured traveltime for all transmitter-receiver pairs $(\alphav_{i}, \betav_{j})$ 
as a vector 
$\tv = \{ t_{1}, \dots, t_{n} \}$, where $n = p \cdot q $. 
Assuming that there are $N$ objects in the velocity model, 
we can group all parameters and define a vector of model parameters, 
$\boldsymbol{\Theta}  = \{ \thetav_{1}, \dots, \thetav_{N} \} $,  
so that the velocity model $V(\thetav_{1}, \dots, \thetav_{N})$ can be
represented simply by $V(\boldsymbol{\Theta})$. 

The predicted traveltime, based on the velocity model parameters
$\boldsymbol{\Theta}$, will be a vector function
$\Tm(\boldsymbol{\Theta}, \mathcal{A}, \mathcal{B})$ representing  the
traveltime between each pair of transmitters and receivers: 
\begingroup\makeatletter\def\f@size{9}\check@mathfonts 
\begin{equation}
\Tm(\boldsymbol{\Theta}, \mathcal{A}, \mathcal{B})  =  (T_{1}(\boldsymbol{\Theta}, \mathcal{A}, \mathcal{B}),
 \dots, T_{n}(\boldsymbol{\Theta}, \mathcal{A}, \mathcal{B})) ,
\end{equation}
\endgroup
where 
\begingroup\makeatletter\def\f@size{9}\check@mathfonts 
\begin{equation}
\label{eqn:travel_time_vector_define}
\left\{ ~\
\begin{matrix}
T_{1}(\boldsymbol{\Theta},  \mathcal{A}, \mathcal{B}) & =  &
\tau^{*}(V(\thetav_{1}, \dots, \thetav_{N}), \alphav_{1}, \betav_{1}) \\
& \vdots & \\
T_{n}(\boldsymbol{\Theta},  \mathcal{A}, \mathcal{B}) & =  &
\tau^{*}(V(\thetav_{1}, \dots, \thetav_{N}), \alphav_{p}, \betav_{q}). \\
\end{matrix}
\right.
\end{equation}
\endgroup

\subsection{Posterior Probability}

To recover model parameters that match the measured traveltime,
we use the quadratic data-fitting error
between the predicted traveltime and the measurements as the cost function:
\begingroup\makeatletter\def\f@size{9}\check@mathfonts 
\begin{equation}
\label{eqn::data_fitting_fn}
 E(\boldsymbol{\Theta})  =  [ \tv - \Tm(\boldsymbol{\Theta},
 \mathcal{A}, \mathcal{B}) ]^{t} \Cm_{t}^{-1}  [ \tv - \Tm(\boldsymbol{\Theta},
 \mathcal{A}, \mathcal{B}) ] 
\end{equation}
\endgroup
where $\Cm_{t}$ is the covariance matrix of measured traveltime. If the
measurement noise is an independent and identically distributed
(i.i.d.) Gaussian random variable, the covariance matrix will be proportional to an identity matrix, 
$\Cm_{t} = \rho \cdot \Id$, and the cost function can be simplified as $ E(\boldsymbol{\Theta})  =  \| \tv - \Tm(\boldsymbol{\Theta},
 \mathcal{A}, \mathcal{B}) \|^{2} $. 
Then, the likelihood function, which is the probability density of the
measurement $\tv$ given the model $\boldsymbol{\Theta}$ will be:
\begingroup\makeatletter\def\f@size{9}\check@mathfonts 
\begin{equation}
  \label{eq:fn_gaussian}
  f(\tv|\boldsymbol{\Theta}) = \frac{1}{Z} \cdot \text{exp}[
   -E (\boldsymbol{\Theta}) / 2 ]
\end{equation}
\endgroup
where $Z$ is a normalization constant. 
We use the Bayesian approach to combine the prior
knowledge and the measurements. So that the posterior probability density
$f(\boldsymbol{\Theta} | \tv)$ is defined
by Bayes rule: 
\begingroup\makeatletter\def\f@size{9}\check@mathfonts 
\begin{align}
  \label{eq:fn_posterioi}
  f(\boldsymbol{\Theta} |\tv) & =   f(\tv |
  \boldsymbol{\Theta} ) \cdot \frac{f( \boldsymbol{\Theta} )}{f(\tv)} \\
  & = k \cdot   f(\tv | \boldsymbol{\Theta} ) \cdot f(
  \boldsymbol{\Theta} )
\end{align}
\endgroup
where $f( \boldsymbol{\Theta} )$ is the prior probability density of model
$\boldsymbol{\Theta} $, and $k$ is a normalization constant. 

We can use maximum a posteriori (MAP) estimation on
(\ref{eq:fn_posterioi}) to estimate a single model, 
but when the measured
traveltime is noisy and the number of measurements is limited,
providing one single model may not be the best way to explain the
data. In this case, it is highly likely that there will be multiple models that reach the
minimum of the cost function. 
In the next section, we use the Hybrid Monte Carlo (HMC) method to
sample the probability density and use the resulting samples to
generate an ``ensemble'' model to explain the data.

\subsection{Proposed algorithm}
\label{subsec:high_contrast_algo}

Because the probability density function is defined by the cost
function, and the cost function of any model can be calculated based on the mismatch between the prediction of traveltime function
and measured data, we can easily evaluate the probability at any point in model parameter space.
However in a high dimensional space, it is not easy to draw samples from a given
probability distribution \cite{mackay2003information}. 
For example, a naive approach would be to divide the
whole parameter space into uniform grids, evaluating the probability
in every point and drawing the samples proportional to the probability.   
However, the number of points to be visited grows exponentially with the dimensions of the space. 
If a velocity model has $3$ objects, and
each object requires $5$ parameters to describe its properties, 
this model will have $15$ parameters. If we divide each parameter into $10$ grids,
we need to evaluate the probability in $10^{15}$ different points, which clearly makes this uniform probing approach impractical. 

The Metropolis method \cite{hastings1970monte} is a widely used approach
to generate samples from a high-dimensional distribution.  
It is an example of Markov chain Monte Carlo method (MCMC), where
samples are generated by random walk and we decide to accept them or reject them 
based on target density. Unlike importance sampling or rejection sampling, it does not suffer
as much from the ``curse of dimensionality''. 
However, due to the random walk behavior in the Metropolis algorithm, it
could take a long time to converge to the target density.
To overcome this problem, we use the Hybrid Monte Carlo (HMC)
algorithm \cite{neal2010mcmc} which explores the parameter space more efficiently and speeds up the sampling process.

\subsubsection{Hamiltonian Dynamics}
  
The Hybrid Monte Carlo is a Metropolis method which simulates Hamiltonian dynamics to draw new samples. 
In HMC, we define a dynamical system where the model parameter vector $\boldsymbol{\Theta}$ is augmented by a 
 momentum vector $\pv$, 
where $\boldsymbol{\Theta}, \pv$ have the same size and $\pv$ is
randomly chosen. 
The total energy  $H(\boldsymbol{\Theta}, \pv)$ of the dynamical system is defined as the sum of
``kinetic energy'' and the ``potential energy'', where the ``potential energy''
is equal to the error function $ E(\boldsymbol{\Theta})$ and  the  ``kinetic energy'' is given by $K(\pv)
= \| \pv \|^{2} / 2$. Thus, $H(\boldsymbol{\Theta}, \pv) =
E(\boldsymbol{\Theta}) + K(\pv) $ and the change of state is determined by the 
Hamiltonian mechanics:  
\begingroup\makeatletter\def\f@size{9}\check@mathfonts 
\begin{subequations}
\label{eqn:HMC_sim_equation}
\begin{align}
\frac{\partial {\boldsymbol{\Theta}} }{\partial t} & =  \pv, \\
\frac{\partial{\pv}}{\partial t} & =  \frac{
  - \partial{E(\boldsymbol{\Theta})}}{\partial{\boldsymbol{\Theta}}}
\end{align}
\end{subequations}
\endgroup

To generate samples for model parameters, we choose a random momentum
and calculate the change of parameters by solving the Hamiltonian dynamics
in  (\ref{eqn:HMC_sim_equation}). 
This process can be viewed as sampling from the joint density 
\begingroup\makeatletter\def\f@size{9}\check@mathfonts 
\begin{align}
  \label{eq:joint_hmc}
  P_{H}(\boldsymbol{\Theta}, \pv) & = \frac{1}{Z_{H}}
  \text{exp}[-H(\boldsymbol{\Theta}, \pv)] \nonumber \\
& = \frac{1}{Z_{H}}
  \text{exp}[-E(\boldsymbol{\Theta})]   \text{exp}[-K(\pv)] 
\end{align}
\endgroup
Because the density is separable, $P_{H}(\boldsymbol{\Theta}, \pv) =
P(\boldsymbol{\Theta})P(\pv)$, we can ignore the momentum
variable and obtain the samples $\boldsymbol{\Theta}^{(t)}$ that are
asymptotically generated from $P(\boldsymbol{\Theta})$.

\subsubsection{Leap Frog Algorithm}
To simulate the Hamiltonian dynamics in discrete time, we use the
``Verlet / Leap-Frog'' algorithm \cite{van1988leap} to maintain the time reversal symmetry. 
Starting from the Taylor's expansion, we can write the discretized
$\boldsymbol{\Theta}_{n}$ as:
\begingroup\makeatletter\def\f@size{9}\check@mathfonts 
\begin{align}
  \label{eq:leap-frog}
  \boldsymbol{\Theta}_{n+1} = \boldsymbol{\Theta}_{n} + \Delta_{t}
  \boldsymbol{\Theta}^{'}_{n} + \frac{1}{2} (\Delta_{t})^2
  \boldsymbol{\Theta}^{''}_{n} + \frac{1}{6}(\Delta_{t})^3
  \boldsymbol{\Theta}^{'''}_{n} + O(\Delta_{t}^4) \nonumber \\
  \boldsymbol{\Theta}_{n-1} = \boldsymbol{\Theta}_{n} - \Delta_{t}
  \boldsymbol{\Theta}^{'}_{n} + \frac{1}{2} (\Delta_{t})^2
  \boldsymbol{\Theta}^{''}_{n} - \frac{1}{6}(\Delta_{t})^3
  \boldsymbol{\Theta}^{'''}_{n} + O(\Delta_{t}^4) \nonumber
\end{align}
\endgroup
By adding them together, we have the following equation:
\begingroup\makeatletter\def\f@size{9}\check@mathfonts 
\begin{equation}
  \label{eq:leap-frog2}
  \boldsymbol{\Theta}_{n+1} = 2\boldsymbol{\Theta}_{n} -
  \boldsymbol{\Theta}_{n-1} + (\Delta_{t})^2
  \boldsymbol{\Theta}^{''}_{n} + O(\Delta_{t}^4) 
\end{equation}
\endgroup
And we define the discrete $\pv_{n}$ and $\boldsymbol{\Theta}^{''}_{n}$ in equation (\ref{eqn:HMC_sim_equation}):
\begingroup\makeatletter\def\f@size{9}\check@mathfonts 
\begin{subequations}
  \label{eq:leap-frog-pv}
  \begin{align}
    \pv_{n - \frac{1}{2}} & =  ( \boldsymbol{\Theta}_{n} -
    \boldsymbol{\Theta}_{n-1} ) / \Delta_{t} \\
    \boldsymbol{\Theta}^{''}_{n} & = ( \pv_{n+ \frac{1}{2}} - \pv_{n -
      \frac{1}{2}} ) / \Delta_{t} 
  \end{align}
\end{subequations}
\endgroup

Substituting (\ref{eq:leap-frog2}) into (\ref{eq:leap-frog-pv}),
it becomes 
\begingroup\makeatletter\def\f@size{9}\check@mathfonts 
\begin{align}
  \label{eq:leap-frog-combine}
  \boldsymbol{\Theta}_{n+1} & = \boldsymbol{\Theta}_{n} + \Delta_{t} (
  \pv_{n - \frac{1}{2}} + \Delta_{t} \boldsymbol{\Theta}^{''}_{n} ) +
  O(\Delta_{t}^4) \nonumber  \\
  & = \boldsymbol{\Theta}_{n} + \Delta_{t} \cdot \pv_{n + \frac{1}{2}}  +   O(\Delta_{t}^4)
\end{align}
\endgroup
Thus we have the Leap-frog algorithm for simulating the discrete Hamiltonian dynamics: 
\begingroup\makeatletter\def\f@size{9}\check@mathfonts 
\begin{subequations}
\begin{align}
  \pv_{n+ \frac{1}{2}} &  = \pv_{n} - \frac{\Delta_{t}}{2} \cdot \left. \frac{\partial
    E(\boldsymbol{\Theta})} {\partial
    \boldsymbol{\Theta}} \right|_{
  \boldsymbol{\Theta}_{n}}  \\ 
  \boldsymbol{\Theta}_{n+1}  & =  \boldsymbol{\Theta}_{n} + 
  \Delta_{t} \cdot \pv_{n + \frac{1}{2}}  \\
  \pv_{n+ 1} & = \pv_{n + \frac{1}{2} } -  \frac{\Delta_{t}}{2} \cdot \left. \frac{\partial
    E(\boldsymbol{\Theta})} {\partial
    \boldsymbol{\Theta}} \right|_{
  \boldsymbol{\Theta}_{n+1}}
\end{align}
\end{subequations}
\endgroup

\subsubsection{Accept / Reject Rules}
After simulating the Hamiltonian dynamics for $n$ leap-frog steps, we
will reach a new state from $\boldsymbol{\Theta}$ to
$\boldsymbol{\Theta^{*}}$. Then we use the Metropolis rule to decide
if we should accept the new state or stay in the current state. 
The new state is accepted with probability:
\begingroup\makeatletter\def\f@size{9}\check@mathfonts 
\begin{equation}
  \label{eq:metro_accept}
  P_{\text{accept}} = min \left( 1,
    \frac{\text{exp}(-H(\boldsymbol{\Theta^{*}},
    \pv^{*}) )} {\text{exp}(-H(\boldsymbol{\Theta}, \pv) ) } \right)
\end{equation}
\endgroup
If the simulation of the Hamiltonian dynamics is perfect, it will
preserve the total energy $H(\boldsymbol{\Theta}, \pv)$ and new state
is always accepted. 
However, due to the finite step size $\Delta t$ the conservation of total energy is not guaranteed.  
Thus, the rejection rule in (\ref{eq:metro_accept}) makes 
sure that the samples are coming from the target joint density
$P_{H}(\boldsymbol{\Theta}, \pv)$.

\begin{algorithm}[htb]   
\begin{algorithmic}

\State {$\boldsymbol{\Theta} = \boldsymbol{\Theta}_{init} $}
\Comment{\small Initialize model parameters}

\For{$l = 1 : L$}               \Comment{ Repeat the simulation for L times }

 \State {$\gv = \nabla E( \boldsymbol{\Theta} ), ~ E= E( \boldsymbol{\Theta} )$}      

   
 \State {$\pv \leftarrow $  randn( size($\boldsymbol{\Theta}$) ) }

 \State {$ H = \pv^{t}\pv/2 + E$ } 

\State{$\boldsymbol{\Theta}_{new} = \boldsymbol{\Theta}, ~\gv_{new} = \gv$} 

\For{$t =  1: T$} \Comment{Use ``leapfrog'' to simulate}

\State {$ \pv = \pv - {\Delta t}/2 \cdot \gv_{new} $ }  

\State 
{$
\boldsymbol{\Theta}_{new} = \boldsymbol{\Theta}_{new} + \Delta t \cdot \pv
$}

\State 
{$ \gv_{new} = \nabla E(\boldsymbol{\Theta}_{new}) $} 

\State 
{$ \pv = \pv - {\Delta t}/2 \cdot \gv_{new}  $}

\EndFor

\State  {$E_{new} = E(\boldsymbol{\Theta}_{new})$} 

\State  {$H_{new} = \pv^{t}\pv /2 + E_{new}$}  

\State 
{$ \Delta H = H_{new} - H$} \Comment{Accept sample by Metropolis rule}

\If{$ \Delta H < 0$}

\State {$\boldsymbol{\Theta} = \boldsymbol{\Theta}_{new}$} 

\Else
\If{$\text{exp}(- \Delta H) \geq \text{rand()}$}

\State {$\boldsymbol{\Theta} = \boldsymbol{\Theta}_{new}$}

\EndIf

\EndIf 

\EndFor

\end{algorithmic}
\caption{Hamiltonian Monte Carlo method}
\label{alg:HMC}
\end{algorithm}

\section{Simulation Results}
\label{sec:sim_results}

To validate our approach, we run simulations on a model that contains
a high velocity structure in the homogeneous background. 
The model size is $100$ m by $160$ m, where the transmitters and
receivers are placed on the boundary with equal spacing. 
The synthetic measured traveltime data is generated by the fast marching
method (FMM) on a grid-based model with cell size $1 \text{m}
\times 1 \text{m}$, where the ground truth velocity distribution has a high 
velocity structure equal to $100 ~\text{m/s}$, with background set to $1~\text{m/s}$. 
We simulate the noisy measurements by adding independent random noise to the
synthetic traveltime, where the value is drawn from a normal
distribution with zero mean and different signal to noise (SNR) level.

In our simulations, we simplify our object model by restricting the ``type of object'' 
parameters. We do so by using rectangles as the only type of fundamental objects to
approximate the structure. The reason is that the distance between two rectangles can be 
calculated analytically from their parameters, and this property significantly reduces the computational cost. 
The initial model is a homogeneous
background with velocity $1 ~\text{m/s}$, and all objects have the same 
parameters - they are located in the center of area-of-interest, 
with same size and orientation. Then we run HMC sampling to sample the 
model parameters. Note that initially these objects exactly overlap with 
each other, and they start to move around and approximate the structure 
through iterations.
In HMC sampling, the gradient of total energy 
is calculated by numerical differentiation. This step is the most computationally expensive part and 
it scales up linearly with the number of objects.

In practice, we need to select suitable values for the the leapfrog step size $\Delta t$
and the number of steps $T$. Generally speaking, having a too large step size will produce 
too much error in simulating Hamiltonian dynamics, thus causing a very low acceptance rate. 
On the other hand, having a too small step size will reduce the exploration rate and need more samples to cover the whole space. Fortunately, by using the leapfrog approach the error in simulating Hamiltonian dynamics 
usually does not increase with the number of steps. However, the number of steps corresponds to the 
``length of trajectory''. If the number of steps is too small HMC will produce highly 
correlated sample points, which means it only explored a nearby region of parameter space. 
In contrast, too many steps may cause the trajectory having the ``turn around'' behavior and 
waste the computational time. For more detail, please refer to Neal \cite{neal2010mcmc}.

In these experiments, we choose the step size and number of steps using an ad-hoc rule. We first run a 
pilot experiment and observe the change of total energy in Hamiltonian dynamics. The empirical rule is 
to choose a suitable step size with limited error in total energy (Ideally, the total energy should be constant 
through the simulation). We choose step size $\Delta t = 0.001$ and number of steps $T=20$ in our experiments.  

To visualize the result, we use a mapping function $f(\boldsymbol{\Theta})$ which maps the
object parameters into the corresponding object shape in spatial domain.  
After we obtain $N$ samples, we can calculate the ``ensemble'' average of models: 
\begingroup\makeatletter\def\f@size{9}\check@mathfonts 
\begin{equation}
  \label{eq:evalue_pdf_map}
\mathcal{M}_{f} = \frac{1}{N}
\sum_{i=1}^{N} 
f(\boldsymbol{\Theta_{i}}) . 
\end{equation}
\endgroup
Because $f(\boldsymbol{\Theta})$ represents the high velocity structure
in spatial domain, $\mathcal{M}_{f}$ can be viewed as the ``probability map'' 
of the high velocity structure. 

\begin{figure}[htb]
  \centering
    \begin{subfloat}[]{
\includegraphics[width=
    0.4\linewidth]{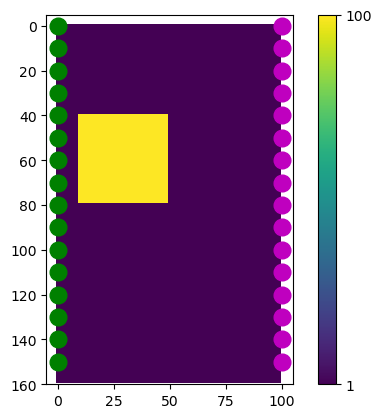}
    }
\end{subfloat}
\begin{subfloat}[]{
   \includegraphics[width=
   0.4\linewidth]{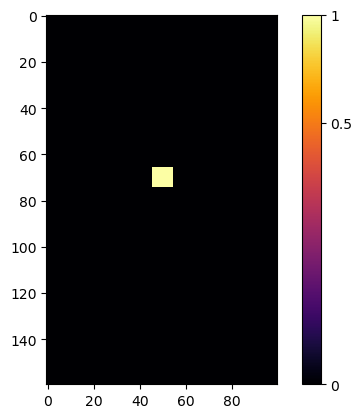}     
} 
\end{subfloat}
\begin{subfloat}[]{
   \includegraphics[width=
   0.4\linewidth]{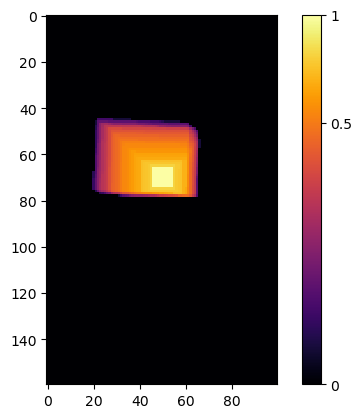}
} 
\end{subfloat}
\begin{subfloat}[]{
   \includegraphics[width=
   0.4\linewidth]{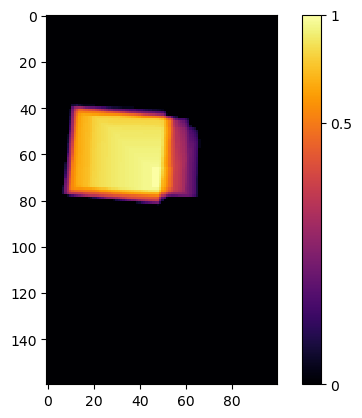} 
} 
\end{subfloat}
  \caption{Results of testing example (a) Ground truth velocity model 
(b) Probability map with $20$ samples (c) Probability map with $100$ samples 
(d) Probability map with $500$ samples.}
  \label{fig:dummy_exp}
\end{figure}

To illustrate the process of HMC sampling, we create a simple experiment where the ground truth velocity model 
has a high velocity square region. 
In this experiment, $16 \times 16$ transmitter-receiver pairs are
placed on the left and right boundary with equal distance $10$ m. 
Transmitter and receiver locations are represented by green and red 
dots, respectively.
The synthetic traveltime data is generated by FMM method with no noise added. 
For reconstruction, we choose the number of object in the model equal to $1$, 
and show the progress of HMC sampling in Fig. \ref{fig:dummy_exp}. 
From the result, we can see the first $20$ samples are still very close to the 
initial model. With $100$ samples, HMC sampling starts to move towards the 
ground truth (high probability region). And with $500$ samples, the HMC 
sampled probability map is very close to our expectation.

\begin{figure}[htb]
  \centering
    \begin{subfloat}[]{
\includegraphics[width=
    0.4\linewidth]{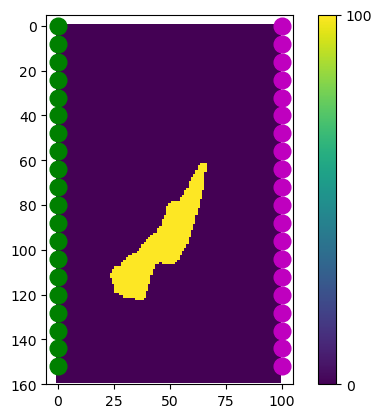}
    }
\end{subfloat}
\begin{subfloat}[]{
   \includegraphics[width=
   0.4\linewidth]{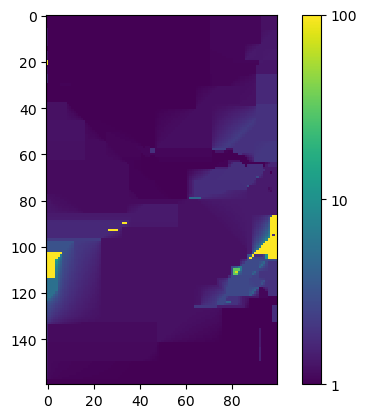} 
} 
\end{subfloat}
\begin{subfloat}[]{
   \includegraphics[width=
   0.4\linewidth]{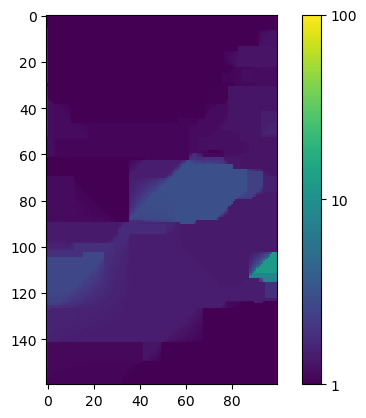} 
} 
\end{subfloat}
\begin{subfloat}[]{
   \includegraphics[width=
   0.4\linewidth]{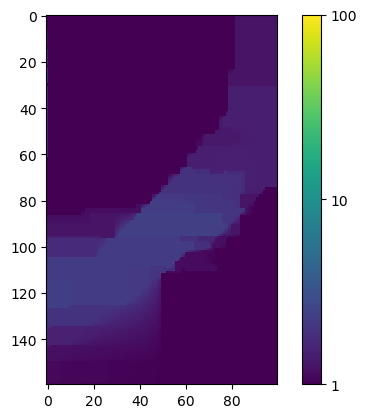} 
} 
\end{subfloat}
  \caption{Results from SIRT (a) Ground truth model (b) Reconstructed velocity model with regularization weight $\alpha = 0.05$ (c) Result with $\alpha = 0.1$ (d) Result with $\alpha = 0.2$} 
  \label{fig:Result_L1}
\end{figure}

For the next experiment, we use a slightly more complicated ground truth model. 
In this experiment, $20 \times 20$ transmitter-receiver pairs are
placed on the left and right boundary and these transceivers are placed with 
equal distance $8$ m. 
We show the ground truth model, with the locations of transmitter as green 
dots and receivers as red dots in Fig. \ref{fig:Result_L1} (a). 
The synthetic traveltime data is generated by FMM method, 
and we run our experiments on both noiseless and noisy measures. 
The noisy measurements are generated by adding independent Gaussian noise 
into each measurements. 
That means each measurement is $\tilde{t_{i}} = t_{i} + \epsilon_{i}$, 
where $t_{i}$ is the traveltime from FMM and $\epsilon_{i}$ is the added noise. 
The added noise is zero mean and independent of each measurement, 
thus $E(\epsilon_{i}) = 0$ and $E(\epsilon_{i} \epsilon_{j}) = 0$. 
The signal-to-noise (SNR) is set to be $10$db, thus the covariance matrix of 
the measurements is a diagonal matrix with $E(\epsilon_{i}^2) = t_{i}^2/10$. 
 
For comparison, we use the bent ray reconstruction with simultaneous iterative reconstruction technique (SIRT), where the models are 
calculated on a regular grid with $100 \times 160$ cells. 
In SIRT, we use the $L_1$ norm of total variation as regularization, 
where the cost function is a combination of data
fitting and weighted regularization term. 
To make the best possible results, we put additional constraints on the value of reconstructed velocity, $1 \leq v \leq 100$ m/s and try different regularization weights. 
Fig. \ref{fig:Result_L1} (b) shows the reconstructed velocity model using noiseless data, with regularization weight equal to $0.05$. The result is very far from the ground truth - it has a few high velocity cells close to the transmitters and receivers. This behavior fits our expectation because the cells close to the transceivers are most sensitive to the change of traveltime. In Fig. \ref{fig:Result_L1} (c)(d) we show the results by increasing the regularization weight to $0.1$ and $0.2$. When we increase the weight of regularization, it favors the result with piece-wise constant velocity model. Although in Fig. \ref{fig:Result_L1} (d) it successfully identifies that the high velocity region is roughly within a $45^{\circ}$ angle, but it provides very poor result for the exact location.  

\begin{figure}[htb]
  \centering
    \begin{subfloat}[]{
\includegraphics[width=
    0.4\linewidth]{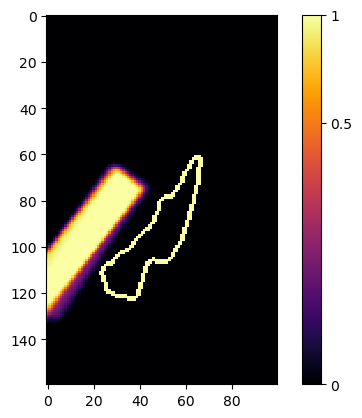}
    }
\end{subfloat}
\begin{subfloat}[]{
   \includegraphics[width=
   0.4\linewidth]{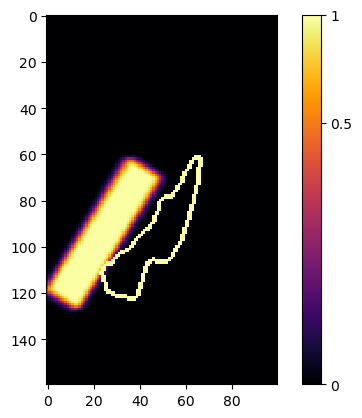}     
} 
\end{subfloat}
\begin{subfloat}[]{
   \includegraphics[width=
   0.4\linewidth]{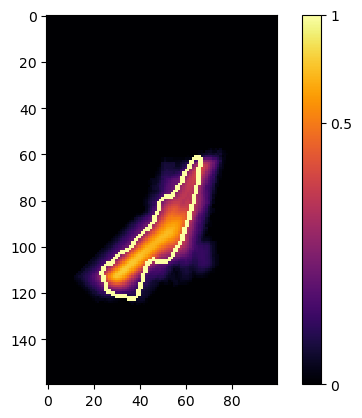}
} 
\end{subfloat}
\begin{subfloat}[]{
   \includegraphics[width=
   0.4\linewidth]{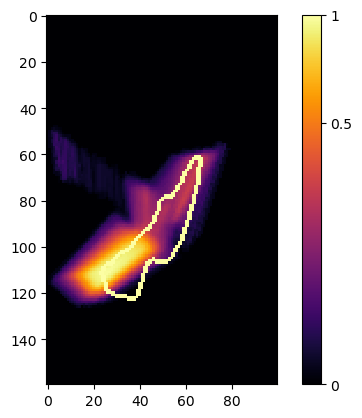}
} 
\end{subfloat}
  \caption{Results from HMC sampling with $5000$ samples (a) One object, noiseless (b) One object, SNR = $10$dB (c) Three objects, noiseless (d) Three objects, SNR = $10$dB}
  \label{fig:Result_HMC}
\end{figure}

\begin{figure}[htb]
  \centering
    \begin{subfloat}[]{
\includegraphics[width=
    0.4\linewidth]{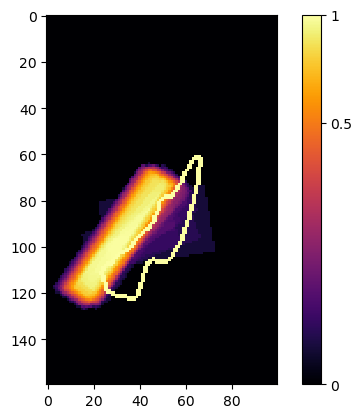}
    }
\end{subfloat}
\begin{subfloat}[]{
   \includegraphics[width=
   0.4\linewidth]{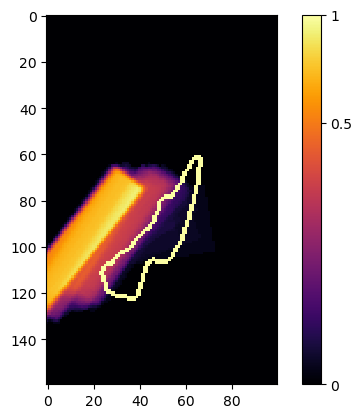} 
} 
\end{subfloat}
\begin{subfloat}[]{
   \includegraphics[width=
   0.4\linewidth]{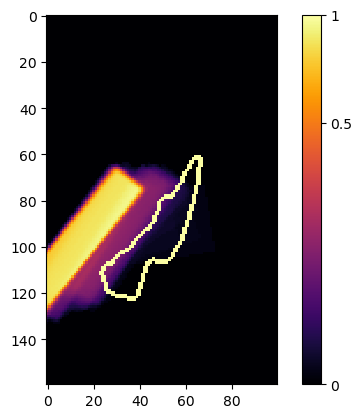} 
} 
\end{subfloat}
\begin{subfloat}[]{
   \includegraphics[width=
   0.4\linewidth]{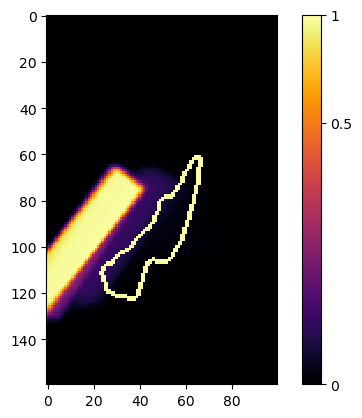}
} 
\end{subfloat}
    \begin{subfloat}[]{
\includegraphics[width=
    0.4\linewidth]{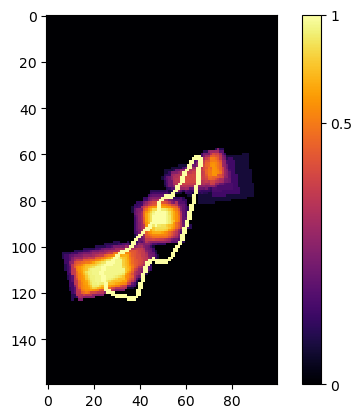}
    }
\end{subfloat}
\begin{subfloat}[]{
   \includegraphics[width=
   0.4\linewidth]{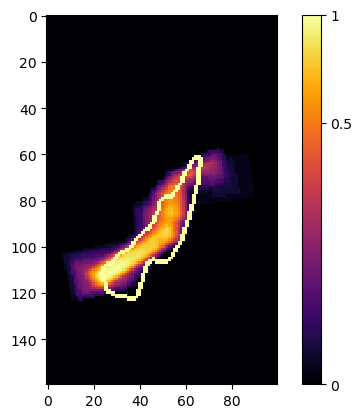} 
} 
\end{subfloat}
\begin{subfloat}[]{
   \includegraphics[width=
   0.4\linewidth]{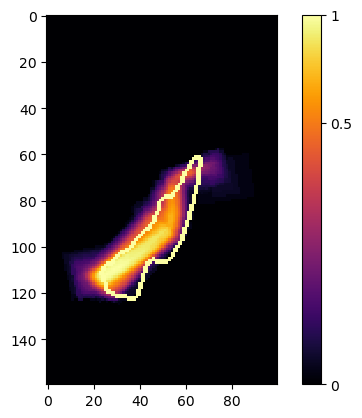} 
} 
\end{subfloat}
\begin{subfloat}[]{
   \includegraphics[width=
   0.4\linewidth]{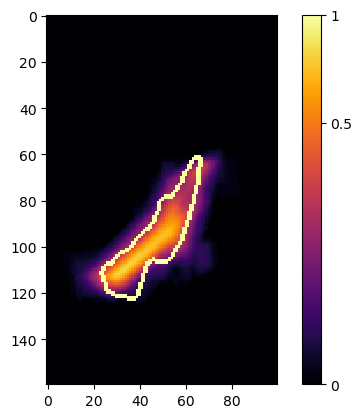}
} 
\end{subfloat}
  \caption{Results from HMC sampling when estimating one object model with 
(a) $100$ samples (b) $500$ samples (c) $1000$ samples (d) $5000$ samples, and 
three objects with (e) $100$ samples (f) $500$ samples (g) $1000$ samples 
(h) $5000$ samples}
  \label{fig:Result_HMC_Samples}
\end{figure}

We show the "probability map'' from our approach in Fig \ref{fig:Result_HMC}. In this experiment, we take $5000$ samples and drop the first $1000$ to avoid the transient state of Markov Chain. For better visualization, we mark the boundary of ground truth model. Fig \ref{fig:Result_HMC} (a)(b) shows that one object model can recover the approximate location and angle of high velocity region, but also has limited ability to model the shape. However we still observe some level of mismatch, especially for the shift in the horizontal axis. This effect is very common in traveltime tomography. 
In our setting, most transmitter-receiver pairs have stronger 
horizontal component. Thus, we expect to have better resolution along the vertical axis 
as compared to the horizontal axis.

Fig. \ref{fig:Result_HMC} (c)(d) shows the result for the 
three object model. Obviously, the three object model provides much better 
approximation for the shape, but it is also more sensitive to noise. 
The reason is that three object model has more degrees of freedom, 
thus it could `over-fit` the data. Also, we want to mention that when running HMC with limited 
samples a more complex model may not provide better results. The reason is that if a 
model has more degree of freedom, it also requires many more samples to converge to the target probability distribution. With limited samples, the Markov Chain may stay in transient states and thus the sampled probability will not reflect the target distribution. We show the results for different number of samples in Fig. \ref{fig:Result_HMC_Samples}, which shows that one object model requires fewer samples to converge. For more detail on this topic, please refer to \cite{raftery1996implementing}.

Note that Fig. \ref{fig:Result_HMC_Samples} (a-d) 
shows that the result converges to a wrong location (horizontal shift from 
ground truth).  Because most transmitter-receiver pairs 
have a stronger horizontal component, we expect to have better resolution along 
the vertical axis as compared to the horizontal axis. 
That means that two models where their objects shifted with respect to each 
other along the horizontal direction could have very similar 
probability. Fig. \ref{fig:Result_HMC_Samples} (a) may give the impression
that the result with few samples is better, but it is 
actually still in a transient period of Markov Chain moving from the initial 
state.

Compared to the result in Fig. \ref{fig:Result_L1}, it is obvious that our approach 
provides a more accurate reconstructed model for the high velocity structure 
in the sparse measurements case. 
The result from SIRT suffers from the ``curse of dimensionality'',
that is, we only have $20 \times 20$ data points but we try to recover $100
\times 160$ unknown variables. 
Although we use total variation as the regularization (which favors a 
piece-wise constant solution), no matter how we change the weight of regularization, 
SIRT still fails to provide a good estimate of the location of high velocity zones.  
The reason is that the data is ``too sparse'' for SIRT. 
For estimating the velocity in a cell, we need at least two trajectory passing through it. 
However, in this experiment the ratio between the number of cells and trajectories is 
$2.5\%$ and most cells have zero or one trajectory through them. 
Thus, SIRT can assign the high velocity cells arbitrary along the trajectory and it usually estimates the high velocity structure to be closed to 
the transceivers, which is a common phantom for SIRT with sparse measurements. 
The reason is that the velocity in these areas is most sensitive to
the change of cost function. Thus, SIRT will favor a
velocity model that changes most near the transmitters or receivers, 
while our approach is more robust when only limited amounts of data are available. 

\section{Conclusion}
\label{sec:conclusion}

The main purpose of this paper is to propose a new approach for the reconstruction of 
velocity models in travel time tomography. 
We use the object-based model to approximate the velocity structure, 
and formulate the reconstruction as a probability sampling problem in parameter space. 
We use the misfit of traveltime to define a potential function, 
and apply HMC to sample the parameter space. 
Typically HMC is considered to be too computationally expensive for tomography problems, but we demonstrate that by using some specific type of fundamental objects 
for high contrast media, the traveltime calculation can be greatly simplified.   
Compared to the conventional fast marching method on the cell, our approach provides 
similar results with much less computations. 
This algorithm makes HMC sampling computationally feasible.

In simulations, our approach successfully finds possible models and assigns 
them appearance probability. 
Especially for limited amounts of data, we show that our algorithm provides 
significantly better results than 
typical iterative ray-based reconstruction on cells. 
The only problem is that even with our fast traveltime calculation, a 
$20 \times 20$ TX-RX, three object model simulation takes about $10$ hours 
to generate $5000$ samples in a Xeon workstation, which has a $8$ cores INTEL E5-2670 CPU.
While SIRT with $L_{1}$ regularization only takes few minutes of computational 
time, our approach is much more computationally expensive. 
However, in our algorithm the calculation of cost function is parallalizable 
and fits into the MapReduce framework.  
Ideally, the simulation time could be reduced to $1/N$ if we have $N$ machines.  
With the progress of cloud platform and parallel programming, we believe computation cost would not be a 
big problem in the future and our approach might be widely accepted.
 
Future work will be to extend our approach to general cases. For example, the object model can be used  
not only for high contrast cases but also for general velocity model. Unfortunately, fast traveltime 
calculations may not be available for general cases. Of course, we can always map the object model into cells 
and use fast marching method on it but the computational overhead makes the HMC sampling less attractive. 
We plan to explore the problem of how to define an object model that has an efficient representation of 
velocity structure, and is also applicable to fast traveltime calculation.

Another topic we would like to explore is that of finding the optimum number of objects in the model. 
In the early development of our algorithm, we did consider to put the number of objects as a parameter in our model. 
Then we realize that increasing the number of objects will provide the ability to achieve lower cost function.
Thus, without any prior information about the number of objects, in HMC sampling the number of objects will keep growing 
and the Markov chain sampling never converges. 
Furthermore, if the number of object grows it implies the number of parameters also grows. 
At this point, it is still unclear how to change the step size and number of steps dynamically 
when the dimension changes. 
We note that recent advances in transdimensional tomography \cite{bodin2012transdimensional} may be useful to quantify the trade-off between the number of objects and 
computational complexity of our approach.
Finally, we mention that the Python implementation of our algorithm is 
publicly available at \texttt{https://github.com/STAC-USC}.

\bibliographystyle{spmpsci}      
\bibliography{thesis_Ref}   

\end{document}